\documentclass[aps,prc,twocolumn,showpacs,groupedaddress]{revtex4}
\usepackage{bm}
\usepackage{amsmath}
\usepackage{epsfig}

\begin{document}
\title{Role of deformation on giant resonances \\
within the QRPA approach and the Gogny force}
\author{S. P\'eru}
\email{sophie.peru-desenfants@cea.fr}
\author{H. Goutte}
\affiliation{
CEA/DAM - Ile de France, Service de Physique Nucl\'eaire, Bruy\`eres-le-Ch\^atel, 91297 Arpajon Cedex, France\\}
\date{Received: date/ Revised version: date}

%\voffset=-3truecm
%\hoffset=-1.truecm

%\textwidth 16cm          %définit la largeur du texte
%\textheight 24cm

%\normalsize
%\topmargin 15pt
%\headsep 10pt
%\tolerance 1000
%\begin{document}           % End of preamble and beginning of text.
%
%\baselineskip 22pt        % définit l'interligne
%\baselineskip 11pt        % définit l'interligne
\def\dspt{\displaystyle}

\begin{abstract}
Fully consistent axially-symmetric-deformed 
 Quasi-particle Random Phase Approximation (QRPA) calculations have been performed, in which the same 
Gogny D1S effective force
has been used for both the Hartree-Fock-Bogolyubov mean field and the QRPA approaches.
Giant resonances calculated in deformed $^{26-28}$Si and  $^{22-24}$Mg nuclei as well 
as in the spherical $^{30}$Si and $^{28}$Mg isotopes are presented.
Theoretical results  for isovector-dipole and isoscalar monopole, quadrupole, and octupole responses
are presented and the impact of the intrinsic nuclear deformation is discussed.
\end{abstract}

\pacs{{21.60.Jz}, {24.30.Cz}, {27.30.+t}}
\maketitle
\section{Introduction}

A major challenge in theoretical nuclear physics is the development of a universal approach able to describe 
the excited states of all nuclear systems with the same accuracy.
From this perspective, models based on the Random-Phase-Approximation (RPA) \cite{thou61} 
are well-suited in rigid nuclei as they describe on the same footing
both individual and collective excited states.
Such approaches have been successfully applied to different nuclear systems. 
However, most of these first calculations were restricted to spherical nuclei with no pairing correlations.
The generalization of these RPA calculations to all nuclear systems requires to treat explicitly
pairing correlations, the intrinsic nuclear deformation and the continuum coupling. 
In practice, it is hard to achieve all these improvements at once.
Choices have been made according to the physics under study.
Pairing correlations have been included to describe excited states of open-shell nuclei, 
and then the Quasi-particle RPA (QRPA) formalism has been used~\cite{RS,blaizot}.
Because of the increase of the number of unstable nuclei experimentally accessible, 
some effort has been made also to treat the continuum coupling in RPA approaches~\cite{shlomo1975,bertsch1975,hama} 
as well as in QRPA ones~\cite{hagino2001,grasso,matsuo,khan,colo}.
Progress has also been made in the development of high order (Q)RPA calculations~\cite{kamerd93,colo2001,kamerd2004}.
Since the work of Gering and Heiss~\cite{ger}, deformed (Q)RPA models have been developed~\cite{yoshi,hagino,rand}.
Let us note that many of these improvements have also been applied on top of relativistic mean field calculations \cite{vret,dario,litvi}.
It is worth pointing out that all these developments should be made along a consistent line:
it has been shown that neglecting parts of the residual interaction in the RPA equations can strongly affect 
predictions~\cite{fayans,agra63,shlomo,epja,tera,tapas}.
Then, more and more calculations are now performed using fully self-consistent (Q)RPA approaches ~\cite{shlomo,tera2006}.

In this paper, a fully consistent axially-symmetric-deformed  
Hartree-Fock-Bogolyubov (HFB)+QRPA approach
using a unique effective force, namely the D1S Gogny interaction~\cite{gog1}, is presented.
This approach can be applied in open or closed-shell nuclei and in spherical or deformed ones.
The present study is the first application of this method and is devoted to giant resonances in the s-d shell nuclei, 
$^{22-28}$Mg and $^{26-30}$Si.
Many experimental results are available for light nuclei
\cite{lisan,bertrand,cseh,PRC13,PRC15,PRC76,YB2002,yb,russe,russe2,YB2007} 
and a recent reanalysis of experimental data in $^{28}$Si shows that giant resonances in these nuclei are still of great interest~\cite{YB2007}.
These Magnesium and Silicon isotopes display different ground state deformations, 
then they are good candidates for studying the impact of intrinsic deformation on giant resonances.
In the past, the shift in energy of isoscalar response for deformed nuclei was estimated with macroscopic models, as reviewed
in section 4.7 of ref \cite{ww}.

In section~\ref{forma}, an outline of the QRPA method is presented.
Practical aspects of our approach are discussed in section \ref{accu}.
Mean field calculations for  $^{22-28}$Mg and $^{26-30}$Si are presented in section~\ref{sec3},
together with results for intrinsic deformations and pairing energies. 
Isovector dipole responses are discussed in section~\ref{secD} 
and isoscalar monopole quadrupole and octupole results in section~\ref{qrpaisos}.
Results are compared with experimental data and systematic laws.
Conclusions are drawn in section~\ref{sec4}.

\section{Formalism}\label{forma}
 
Approaches based on the RPA
have been found to be successful in explaining low-lying multipole vibrations and giant resonances.
The QRPA extension, which allows one to treat on the same footing particle-hole~(ph), particle-particle~(pp)
and hole-hole~(hh) excitations is well suited for both closed-shell and open-shell nuclei.
In particular QRPA approaches on top of HFB calculations describe collective  vibrations of 
the nuclear mean field and pairing field
together with  induced ground state correlations.

In the QRPA approach, the quasi-boson operators $\theta^+_n$ representing nuclear excitations are defined as:
\begin{equation}\displaystyle
\theta^+_n=\frac{1}{2}\sum_{ij} \left( X_n^{ij} \eta^+_i \eta^+_j -Y_n^{ij}\eta_j \eta_i\right),
\label{phonon}\end{equation}
where $\eta^+_i$ and $\eta_i$ are  quasi-particle (qp)  creation and annihilation operators, respectively. 
 $X_n$ and $Y_n$ are the amplitudes of  the two quasi-particles (2-qp) excitations.
The QRPA ground state $\vert \tilde{0}>$ and the QRPA excited states $\vert n>$ satisfy the equations:
\begin{equation}\displaystyle
\theta^+_n  \vert \tilde{0}>=\vert n>,\ \ \theta_n  \vert \tilde{0}>=0 .
\label{qrpast}\end{equation}

QRPA equations are obtained through the second derivative of the energy functional built with an effective interaction, 
in the present work the Gogny force D1S~\cite{gog1}.
QRPA equations  can be converted
into a matrix equation:
\begin{equation}\label{equaref}
\left(\begin{array}{cc} { A}& { B}\\{ B}&{ A} \end{array}  \right)
 \left(\begin{array}{c}{X_n}\\ {Y_n}\end{array}\right)
= \omega_n \left(\begin{array}{c}X_n\\-Y_n\end{array}\right),
\end{equation}
where  $\omega_n$ are the energies of the QRPA excited  states $\vert n>$.
In Eq.~(\ref{equaref}), the  matrices A and B are real and symmetric since 
the time reversal symmetry $T$ and the $T\Pi_2$ symmetry, where $\Pi_2$
 is the reflection with respect to the (x0z) plane, are assumed.
 A and B are built using the same Gogny D1S effective force in all ph, pp and hh channels.

To obtain the QRPA energies $\omega_n$ and the amplitudes $X_n$ and $Y_n$, Eq.~(\ref{equaref}) is rewritten as:
\begin{equation} \dspt
\left( A-B\right)\left( A+B\right)\left(X_n +Y_n \right) 
\dspt = \omega_n^2\left(X_n+Y_n\right).
\label{e1}
\end{equation}
Introducing the  matrix $S$ which diagonalizes $\left(A-B\right)$ with $D=S^{-1}\left(A-B\right)S$,
 Eq. (\ref{e1}) can be rewritten
\begin{eqnarray} \nonumber\label{matab}
D^{1/2}S^{-1}\left(A+B\right)S D^{1/2}
D^{-1/2}S^{-1}\left(X_n+Y_n\right) \\
=\omega^2 D^{-1/2}S^{-1}\left(X_n+Y_n\right),
\end{eqnarray}
i. e. as the eigenvalue equation:
\begin{equation}\label{eigen}
C V_n= \omega^2_n V_n,
\end{equation}
where  $C$ and $V_n$ are defined as:
\begin{equation}\dspt
C= D^{1/2} S^{-1}\left(A+B\right) S D^{1/2},
\label{cmat}
\end{equation}
and
\begin{equation}\dspt
V_n= D^{-1/2} S^{-1}\left(X_n+Y_n\right).
\label{vectV}
\end{equation}
Let us note that since the matrices A and B are real and symmetric, S can be chosen orthogonal ($S^{-1}=S^{T}$). 
Consequently C is a symmetric matrix.
 
The QRPA energies $\omega_n$ are then calculated as the positive square-root of the eigenvalues $\omega^2_n$ of $C$.
Using the same procedure for 
$$\left(A+B\right)\left(A-B\right)\left(X_n-Y_n\right)=\omega_n^2\left(X_n-Y_n\right),$$ similar to Eq.~(\ref{e1}),
and the orthonormalization relation
${X_n}^T X_{n'}-{Y_n}^T Y_{n'}= \delta_{nn'}$
the QRPA amplitudes $X_{ n}$ and $Y_{ n}$ are then obtained as:
\begin{equation}\dspt
\left\{ \begin{array}{lcl}\dspt
X_n & = & \dspt\frac{1}{2}\left[ \frac{D^{1/2}}{\sqrt \omega_n} +D^{-1/2} \sqrt \omega_n\right] V_n \\
\\
Y_n & = & \dspt\frac{1}{2}\left[ \frac{D^{1/2}}{\sqrt\omega_n} -D^{-1/2} \sqrt\omega_n\right] V_n.
\end{array} \right.\label{xy}
\end{equation}
Such a procedure, used to solve the QRPA equations (Eq.~(\ref{equaref})), has the advantage of reducing 
the size of the matrix to be diagonalized by half. The condition of positive-definiteness of at least one of the matrices $A+B$, $A-B$
can be relaxed using a generalized Cholesky decomposition \cite{papa}.

In the present approach, the matrix equations are solved in the finite space of N 2-qp excitations.
Then, $\left(A-B\right)$ and $C$ are N$\times$N matrices and are diagonalized using standard routines giving all 
eigenvectors and  eigenvalues.
Let us note that $\omega^2_n$, the solutions of Eq.~(\ref{eigen}), can be negative and then lead 
to imaginary $\omega_n$ values.
Such complex eigenvalues of the QRPA matrix represent unstable collective modes.
However, as discussed in Ref.~\cite{thou61}, it is always possible to find a solution of the HFB equations 
which makes these collective modes stable and consequently ensures that they have real frequencies.

In the present work, axially-symmetric-deformed~(ASD) HFB calculations 
are performed in even-even nuclei imposing some symmetries,
 $T$, $T\Pi_2$ and also axial and left-right symmetry.
Then the projection K of the angular momentum on the symmetry axis
 and the parity $\pi$ are good quantum numbers.
Consequently ASD-QRPA calculations can be performed separately in each $K^{\pi}$ block.
In the following, the ASD-QRPA states $\vert n\rangle$ of Eq.(\ref{qrpast}) will be labeled with the angular  momentum projection $K$ 
and rewritten as $\vert n K \rangle$. Time-reversed states  will be written $\vert\overline{n K}\rangle$.

States $\vert JM K \rangle_n$ of good angular momentum J are obtained projecting 
the intrinsic excitations $\vert n K \rangle$ according to:
\begin{eqnarray}\nonumber
\vert JM K \rangle_n & \dspt =\frac{\sqrt{2J+1}}{4\pi}\int d\Omega\left( {D^J_{MK}}^*\left(\Omega\right)
 R\left(\Omega\right)
\vert n K \rangle \right.\\
& \left.+ \left(-\right)^{J-K}
{D^J_{M-K}}^*\left(\Omega\right) R\left(\Omega\right)  \vert\overline{nK}
\rangle\right)\label{e2JM},
\end{eqnarray}
where $\Omega$ are Euler angles, $D^J_{MK}$ Wigner rotation matrix elements, and $ R\left(\Omega\right)$ 
 the three dimensional rotation operator.

In the QRPA approach all quantities are calculated relative to their values in the ground state.
The response of the system to an external field, such as a multipole operator $\hat{Q}_{\lambda\mu}$,
is obtained from $\vert \langle~00 0\vert\hat{Q}_{\lambda\mu} \vert JM K \rangle\vert ^2$, 
i.e. the square of the transition matrix elements  between projected  $\vert JM K \rangle$ states of Eq.(\ref{e2JM}) 
and the projected QRPA ground state $\vert 000\rangle$.
Using Eq.(\ref{e2JM}), one gets:
\begin{equation}\displaystyle
\langle 000\vert\hat{Q}_{\lambda\mu} \vert JM K \rangle_n
=\sum_{\nu,\vert \nu\vert \leq J} \Gamma_{JK} \langle \tilde{0}\vert \hat{Q}_{J\nu}\vert n K \rangle\delta_{K\nu} \ ,
\end{equation}
where  $\vert \tilde{0}\rangle$ is the ASD-QRPA vacuum of Eq.(\ref{qrpast}) and 
the $\Gamma_{JK}$ coefficients are obtained from straightforward algebraic calculations.

In an axially-symmetric-deformed nuclear system, the response function of  a given  J$^\pi$
contains different $K^\pi =0^\pi,\pm 1^\pi,...\pm J^\pi$ components.
In spherical nuclei, all these components are degenerated in energy,
then the response functions associated to any multipolarity can be obtained from $K^\pi=0^{\pm}$ results only. 
This property has been used to check the newly-built ASD-QRPA computer code.

The collectivity of each QRPA state is obtained through its contribution
 to the  total Energy-Weighted-Sum-Rule (EWSR)
$M_1(Q_{\lambda\mu})=\sum_{n=1}^{N} {\omega}_n | \langle 000\vert \hat{Q}_{\lambda\mu}\vert JMK\rangle_n|^2$.
The  mean energy   of the QRPA resonances in a given energy interval 
 $\left[E_{min},E_{max}\right]$ is calculated as:
\begin{equation}\label{meanvalue}
<E>_{\left[E_{min},E_{max}\right]}=\frac{M_1\left[E_{min},E_{max}\right]}
{M_0\left[E_{min},E_{max}\right]},
\end{equation}
where the moments  $M_\lambda$ are
\begin{equation} \dspt
M_{\lambda}\left[E_{min},E_{max}\right]=\sum 
_{n} 
\omega ^{\lambda}_n 
| \langle 000\vert \hat{Q}_{\lambda\mu}\vert JMK\rangle_n|^2, 
\label{moment}
\end{equation}
with   $E_{min}\le \omega_n \le E_{max}$.

\section{Calculation aspects}\label{accu}

Before presenting the results on Magnesium and Silicon isotopes obtained with the above formalism,
let us discuss the practical aspects of the present calculations.

First, in the present study, the HFB equations are solved in a finite Harmonic Oscillator (HO) basis.
Consequently the positive energy continuum is discretized.

Second,
since, as discussed by Terasaki et al~\cite{tera} 
 QRPA consistency is much affected by inaccurate HFB results,
the HFB equations have been solved with a very high degree of accuracy and 
extremely well converged quasi-particle states are used in QRPA calculations.

Third, all HFB quasi-particle states are used to generate the 2-qp excitation set (Eq.(\ref{phonon})).
This means that no cut in energy or in occupation probabilities is introduced.
Our model space, includes 9 HO major shells.
This is a large enough space for nuclei studied here.
We checked that the calculated EWSR is altered by less than 2\%
when going from 9 to 11 HO major shells.

Fourth, as already emphasized, the omission in the QRPA calculations of parts of the two-body force which are included
in the HFB mean field and pairing field may lead to large errors in strength functions.
Such a lack of consistency is often responsible for a redistribution of the strength~\cite{fayans,agra63,shlomo,epja,tera,tapas}
and a strong spurious state mixing ~\cite{shlomo}.
In the present calculations, the same effective Gogny D1S nucleon-nucleon force is used
to solve both the HFB and the QRPA equations in all ph, pp and hh channels.
However, since the Coulomb exchange field is not taken into account in HFB, 
the corresponding QRPA terms have been set to zero in order to maintain consistency.

Finally, as RPA-type approaches restore broken symmetries, violations of conservation laws by the HFB wave function 
lead to spurious states. Spurious states are expected to have zero energy eigenvalues in the (Q)RPA spectrum \cite{thou61}.
In the present axially-symmetric-deformed QRPA calculations there are seven spurious states.
Except for two of them related to particle number violation, spurious states 
differ from one another since no two of them have the same $K^\pi$ quantum numbers: 
$K^\pi=0^-$, $\pm 1^-$ for translational ones,
$K^\pi=\pm 1^+$ for rotational ones,
and $K^\pi=0^+$ for particle number ones.
Since our calculations have been performed along a consistent line, with
{\it i}) the same interaction in HFB and QRPA calculations, 
{\it ii}) highly converged HFB calculations,
{\it iii})  no cut-off in single particle excitations,
and {\it iv}) a large enough HO basis,
 these spurious states are easily identified with large values of $X\simeq-Y$ and they are found to be well separated from physical excitations. 
They are found at 0.001~keV energy for $K^\pi=0^+$, around 500 keV energy for $K^\pi=\pm 1^+$ and less than 1.8~MeV for $K^\pi=0^-,\pm 1^-$.
The spurious modes relative to translational invariance are more sensitive to the finite size of the HO basis,
which prevents us from fully restoring translational invariance.
In order to put all these spurious modes at zero energy, renormalization factors could have been introduced.
However, as studied in Ref.~\cite{hagino}, renormalization factors associated to different spurious modes 
would differ from each other and no universal choice exists.
A corrected dipole operator could also be used but, as mentioned in Ref.~\cite{tera}
responses for the corrected and uncorrected operators are indistinguishable in the case of fully consistent HFB+QRPA calculations.
In view of this, the results presented below have been obtained
with uncorrected operators and no renormalization factors has been included.

\section{HFB results for  $^{22-28}$M$g$ and $^{26-30}$S$i$}\label{sec3}
\begin{figure*}
%\begin{center}
\centerline{\hbox{\psfig{file=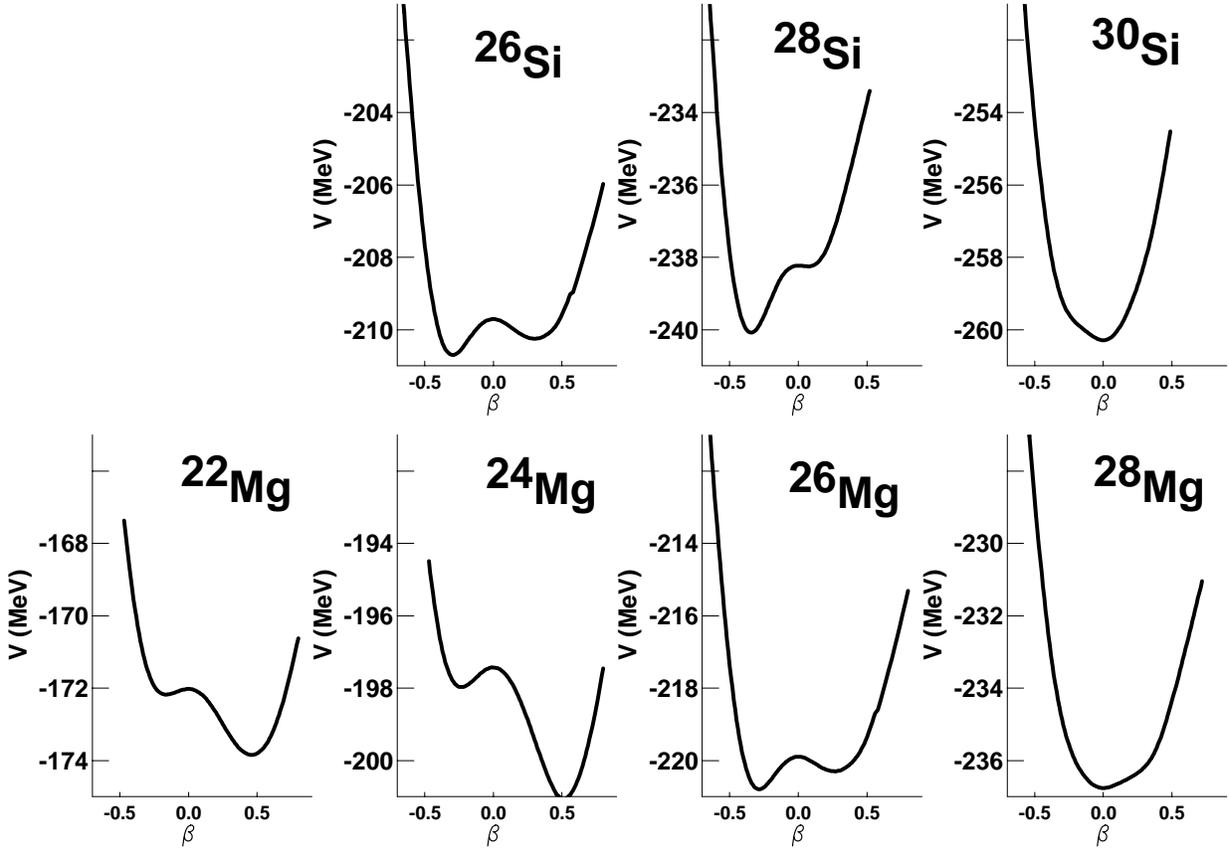,width=18cm,angle=0}}}
\caption{HFB potential energy curves V as functions of the axial deformation parameter $\beta$ 
for $^{26-30}$Si and $^{22-28}$Mg.}
\label{axcurve}
%\end{center}
\end{figure*}

Constrained ASD-HFB calculations have been performed in $^{26-30}$Si and $^{22-28}$Mg nuclei.
Potential energy curves for the seven nuclei are plotted in Fig.~\ref{axcurve}
as functions of the axial
deformation parameter $\beta$:

\begin{equation}
\dspt
\beta =\sqrt{\frac{\pi}{5}} \frac{q_{20}}{A R^2},
\label{beta}
\end{equation}
where $q_{20}$ is the mean value of the axial quadrupole operator $\hat{Q}_{20}\ =\sum_{i=1}^A\sqrt{16\pi/5}\ r^2 (i) { Y}_{20}(i)$,
 and  $R^2\ =\ 3/5 \left( 1.2\ A^{1/3}\right) ^2\ {fm}^2$.

\begin{figure*}
\begin{center}
\centerline{\hbox{\psfig{file=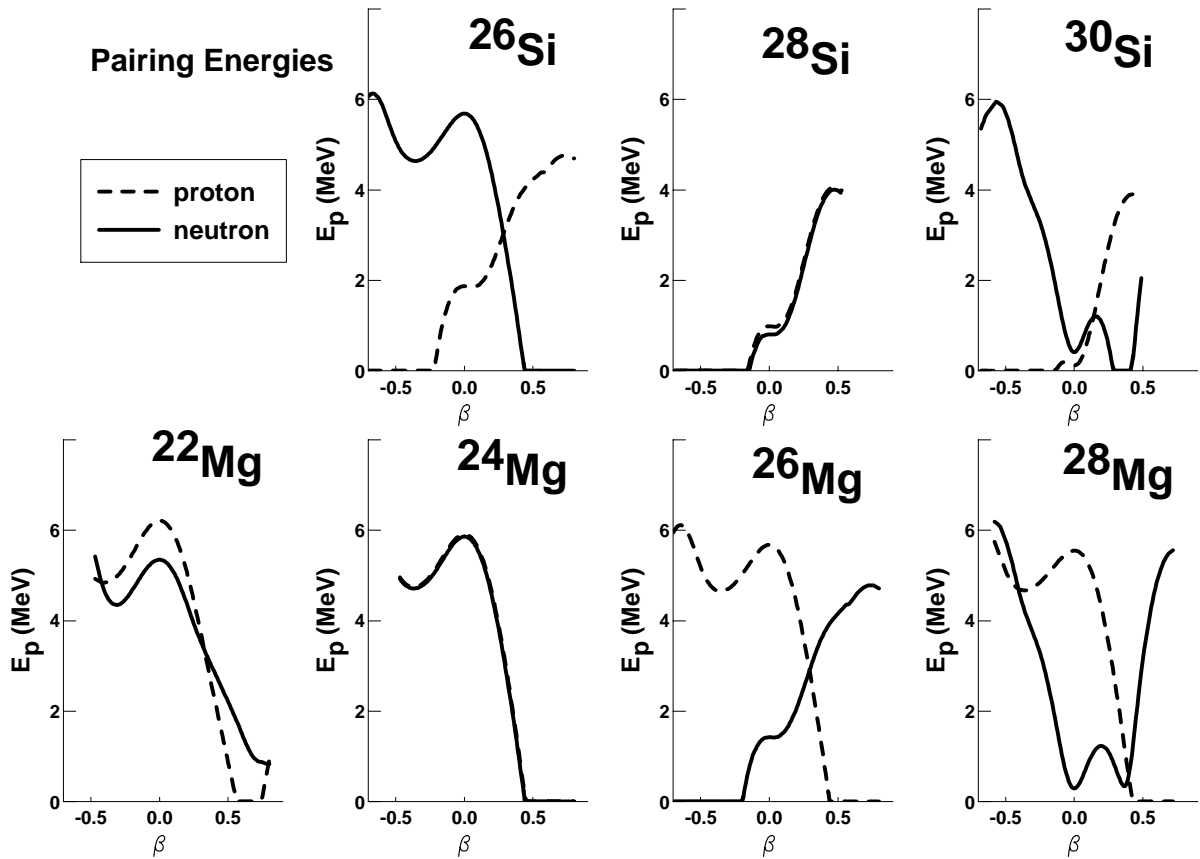,width=18cm,angle=0}}}
\caption{Neutron (full lines) and proton (dashed lines) pairing energies E$_p$ 
 as functions of axial deformation parameter $\beta$ in $^{26-30}$Si and $^{22-28}$Mg.}
\label{pairener}
\end{center}
\end{figure*}

We observe that
{\it i)} the N = 16 nuclei, $^{30}$Si and $^{28}$Mg, are found spherical,
{\it ii)} the Z~=~12 nuclei, $^{22}$Mg and
$^{24}$Mg, are strongly prolate with $\beta$~=~0.47 and 0.52, respectively
and {\it iii)} Z and/or N= 14 ones, $^{26}$Si, $^{28}$Si and $^{26}$Mg,
 present large oblate deformations with $\beta\ \simeq$ -0.3.
In all these nuclei, except spherical ones, the potential energy curves exhibit a secondary minimum as function of axial deformation.
HFB calculations in triaxial symmetry show that these minima are 
connected to the ground state through triaxial deformations.
In these light nuclei, only two protons (or two neutrons) radically change
the deformation of the HFB ground state: this 
is of course a consequence of shell effects, which are also responsible for the vanishing of the pairing energy in specific deformation ranges.
In open shell systems the pairing energy take on a large value. This value is smaller when there is a significant single particle gap above the 
Fermi level. The cancelling of the pairing energy is associated with closed shell systems \cite{n2028}.
Here the pairing energy is defined as
$E_p = \frac{1}{2} Tr \left(\Delta \kappa\right)$ and is calculated separately for neutrons and protons, with $\Delta$ the pairing field  
and $\kappa$  the abnormal density~\cite{n2028}.
In Fig.~\ref{pairener}, proton and neutron pairing energies are plotted as functions of $\beta$
in dashed and solid lines, respectively.
These pairing energies range from 0 to 6 MeV.
The most striking feature of Fig.~\ref{pairener} is that a strong competition is found 
between the Z,~N~=~12 prolate sub-shells and the Z,~N~=~14 oblate ones.
In $^{24}_{12}$Mg$_{12}$ and $^{28}_{14}$Si$_{14}$ nuclei, proton and neutron pairing energies
are canceling at the same deformation:  prolate  for $^{24}$Mg and oblate for $^{28}$Si.
On the contrary in $^{26}$Si and $^{26}$Mg proton and neutron pairing energies do not vanish at the same deformation.
Corresponding  potential energy curves of $^{26}$Si and $^{26}$Mg (Fig.~\ref{axcurve}) present two minima, one for each sub-shell. 
These minima are close in energy, in contrast to those in $^{28}$Si and $^{24}$Mg which are separated
by more than  2 MeV.
In the N~=~16 $^{30}$Si and $^{28}$Mg nuclei, the neutron pairing energy almost vanishes for $\beta \simeq$~0.3, and for $\beta$~=~0
and spherical shapes are stabilized.
The N~=~10 neutron pairing in $^{22}$Mg presents a minimum at very large deformation.

In the following, QRPA calculations are performed using the quasi-particle states associated with the HFB solutions with minimum
energy in all nuclei. These solutions are
either prolate ($^{22-24}$Mg) or oblate ($^{26-28}$Si, $^{26}$Mg) or spherical ($^{28}$Mg, $^{30}$Si).
In $^{24}$Mg and $^{28}$Si, since the total pairing energy is zero
in the HFB ground state, QRPA calculations reduce to conventional RPA ones, where elementary excitations are of particle-hole type. First results
for these two nuclei have been presented previously in a conference proceeding~\cite{comex2}.
In the other nuclei under study, the full QRPA formalism is necessary as either proton 
or neutron pairing correlations are present in the HFB ground state.

\section{Dipole response}\label{secD}
\begin{figure*}
\begin{center}
\centerline{\hbox{\psfig{file=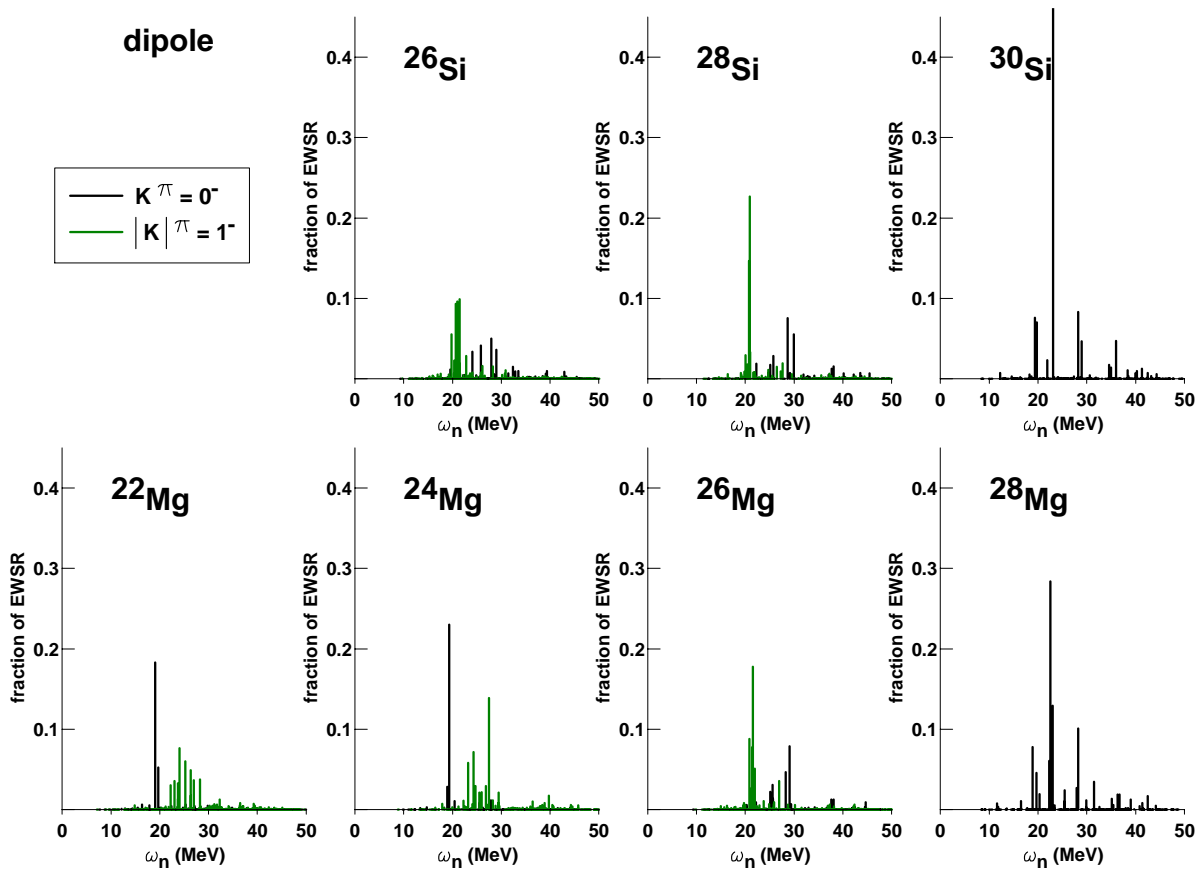,width=18cm,angle=0}}}
\caption{(Color online) Fractions of the isovector dipole EWSR in $^{22-28}$Mg and $^{26-30}$Si.
 In deformed nuclei, K$^\pi=0^-$  and   $\vert$K~$\vert ^\pi=1^-$ components are indicated in black and green, respectively.}
\label{fig5}
\end{center}
\end{figure*}

As mentioned in \cite{ww}, the first evidence of giant-resonance phenomena in nuclei was obtained in 1937 \cite{BOT37}, and
the notion of a dipole oscillation of the nucleus was recognized by Migdal in 1944 \cite{MIG44}.
The results of many years of experimental work on the Isovector Giant Dipole Resonance (IVGDR) have been summarized and reviewed
in several papers 
\cite{BER75,BER77b,DIE88}:
in light nuclei the strength distribution is fragmented into several components,
whereas it is separated into two major components in deformed nuclei.
QRPA responses to the isovector electric dipole operator
 $\hat{Q}_{10}=\sum_{i=1}^{Z} r_i Y_{10}\left( r_i\right)-\sum_{i=1}^{N} r_i  Y_{10}\left( r_i\right)$ 
 are given in Fig. \ref{fig5} as fractions of the EWSR.
The K~$^\pi=0^-$, $\vert$K$\vert ^\pi$~=~$1^-$ components are plotted in black and green, respectively.
In the two spherical nuclei $^{28}$Mg and $^{30}$Si, only the K$^\pi=0^-$ component is drawn
as  $\vert$K~$\vert ^\pi=1^-$ states are  degenerated with K$^\pi=0^-$ ones.
 In all nuclei, the predicted strength is spread over a wide energy range up to 50 MeV, 
 such a broad width being consistent with the fact that 
these nuclei lie in the middle of the s-d shell.
Negligible strength is found at low energy which is in agreement with the fact that pygmy states are observed only in very exotic nuclei.
In spherical nuclei $^{28}$Mg and $^{30}$Si, responses  are similar:
a highly collective state is found at 23~MeV and two less collective components around 20~MeV and 25~MeV.
In deformed nuclei, the strength is found fragmented and splits up into mainly two components,
 as expected.
In this study, the two components correspond to two different angular momentum projections K.
 More precisely, in the prolate  $^{22-24}$Mg nuclei  the lower energy peak is associated to K~=~0 states
 whereas the higher one corresponds to $\vert $K$\vert$~=~J~=~1 states.
The same features are found in the oblate nuclei $^{26-30}$Si and $^{26}$Mg,
with lower energy parts composed of $\vert $K$\vert $~=~J~=~1 states
and higher ones composed of K~=~0 states.
The energy split between K~=~0 and K~=~$\pm$1 states is related to the intrinsic deformation 
of the ground state with an opposite hierarchy for prolate  and oblate shapes.
In prolate nuclei the symmetry axis is the long axis and the
$\vert $K$\vert $~=~J component corresponds 
 to the alignment of the angular momentum along this axis,
 whereas
 for oblate nuclei  the symmetry axis is the shorter one and  the K~=~0 component corresponds 
 to the alignment of the angular momentum along the long axis.
Such a splitting of giant resonance into two groups of transitions,
corresponding to vibrations along the major and minor axis of a deformed spheroid was
already predicted in 1958 by Okamoto~\cite{oka} and Danos~\cite{danos} in a long range correlation model. 
Such an effect was used in order to explain part of the broadening of the resonance widths in non spherical nuclei.

From another theoretical point of view, Bassichis at al.~\cite{bass} have explained 
the GDR splitting in $^{24}$Mg as due to configuration splitting:
the high energy  region (22 - 27 MeV) is formed by 1p$\rightarrow$ (1d-2s) transitions, whereas the lower one
(13 - 22  MeV)  contains  mainly  (1d-2s)~$\rightarrow$~(1f-2p) transitions.
Our results are compatible with this analysis:
major components of the  K$^\pi=0^-$ QRPA state at 19.35~MeV are 1d5/2~$\rightarrow$~1f7/2 transitions
with small contributions of 2p3/2~$\rightarrow$~2s1/2 transitions, whereas
major components of the  $\vert$K$\vert ^\pi=1^-$  QRPA state at 27.44~MeV are  1p3/2$\rightarrow$2s1/2 transitions
with non-negligible components from 1d5/2$\rightarrow$2f5/2 ones.

By comparing our results with experimental data~\cite{russe},
a relatively  good agreement is found in $^{24-26}$Mg.
In Ref.~\cite{russe2} the general structure of the $^{24}$Mg IVGDR is detailed:
a first group of resonances at 18~-~21~MeV forms the main peak of the GDR and a second  one  is a broad plateau
at 21~-~27~MeV with structures at 23~MeV and 25~MeV. The width of the GDR is found to be 9~$\pm$~1~MeV, 
in fair agreement with our results.

As mentioned in Ref.~\cite{ww}, a systematic law can be used to estimate the mean energy 
of the IVGDR,
$\dspt E\left( IVGDR \right)$~=~31.2~$A^{-\frac{1}{3}}$~+~20.6~$A^{-\frac{1}{6}}$~MeV.
Systematic values are compared to our results in Table \ref{tabD}. 
Theoretical values of the IVGDR are obtained by integrating the strength from 15 to 35~MeV.
\begin{table}
\caption{Theoretical (present work) and systematic mean energy values of the IVDGR (in~MeV).}
%\begin{ruledtabular}
\begin{tabular}{ ccc}
\hline\hline
&$E(GDR)$ ~MeV 
$$\\
&th.&Syst.\\
\hline
$^{22}$Mg& 23.09 &23.4   \\
$^{24}$Mg& 23.18 & 22.9  \\
$^{26}$Mg& 23.13 & 22.5  \\
$^{28}$Mg& 23.03 &  22.1 \\
$^{26}$Si& 22.91 & 22.5  \\
$^{28}$Si& 22.80 &  22.1 \\
$^{30}$Si& 23.17 &  21.7 \\
\end{tabular}\label{tabD}
%\end{ruledtabular}
\end{table} 
They are found roughly constant irrespective of the mass number, contrary to the values obtained from the systematics which
vary from 21.7~MeV ($^{30}$Si) to 23.4~MeV ($^{22}$Mg).
In our calculations, the resonance energy is found almost independent of the deformation of the nucleus, 
as already predicted by Danos \cite{danos}.
In all seven nuclei the IVGDR is predicted to exhaust around 85$\%$ of the EWSR.
In prolate $^{22-24}$Mg nuclei, high energy components dominated 
by $\vert$K$\vert ^\pi=1^-$ states are found to exhaust 
 54$\%$ of the EWSR. Low energy components, dominated by  K~=~0 states, exhaust 29$\%$ of the EWSR.
In the three oblate nuclei, high energy components, dominated by  K~=~0 states, exhaust around 35$\%$ of the EWSR
and low energy ones around 50$\%$.
Let us note that in a systematic study~\cite{goriely} Goriely and Khan have shown that an equally distributed strength 
between low energy and high energy components
give optimal agreement with experimental E1 transitions.

\section{Isoscalar multipole responses}\label{qrpaisos}
\subsection{Monopole response}\label{secM}

\begin{figure*}
\begin{center}
\centerline{\hbox{\psfig{file=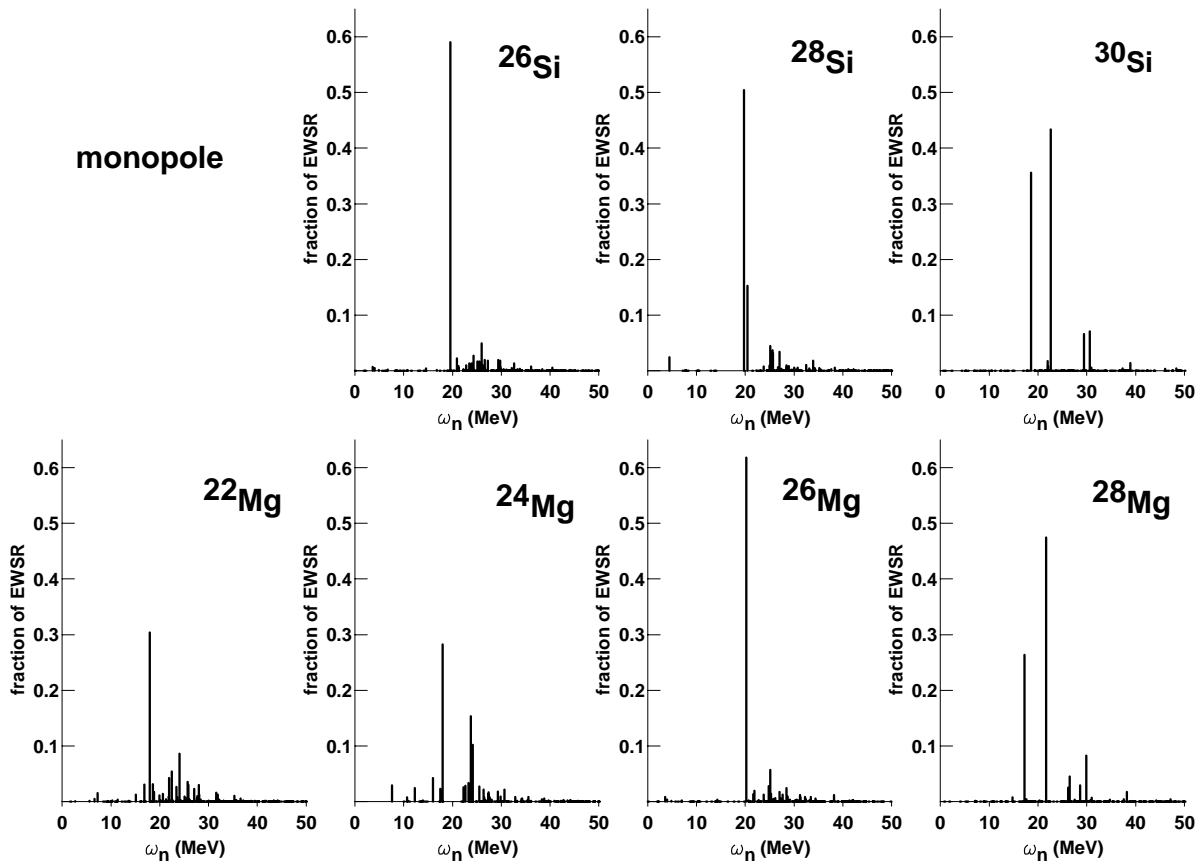,width=18cm,angle=0}}}
\caption{Fractions of the monopole EWSR in $^{22-28}$Mg and $^{26-30}$Si.}
\label{fig4}
\end{center}
\end{figure*}

Fig. \ref{fig4} displays monopole responses of $^{22-28}$Mg, and $^{26-30}$Si
 as fractions of the EWSR.
The most striking feature is that the seven nuclei under study present features that can be separated 
in three classes depending on the deformation of their ground states.

First, the responses of the three oblate nuclei $^{26}$Mg and $^{26-28}$Si display a prominent peak around 20 MeV and a broad
component at a higher energy exhausting less than 30$\%$ of the EWSR.
The low energy part of the resonance exhausts  65$\%$, 62$\%$ and 66$\%$ of the EWSR in $^{26}$Mg, $^{26}$Si and $^{28}$Si, respectively.
The experimental response of the oblate $^{28}$Si \cite{YB2002} displaying a sharp peak close to 18~MeV 
and a large fragmentation up to 35~MeV, is qualitatively reproduced by our results.

Second, the responses of the two prolate nuclei, namely $^{22}$Mg and $^{24}$Mg,
display a two-peaked structure around 18~MeV and 25~MeV,
 the high energy one being very  fragmented.
The low energy component is found to  exhaust 40$\%$ ($^{22}$Mg) and 35$\%$ ($^{24}$Mg) of the EWSR,
and the high energy broad component 44$\%$ ($^{22}$Mg) and 48$\%$ ($^{24}$Mg).

Third, only a few discrete levels emerge in spherical 
 $^{28}$Mg and $^{30}$Si nuclei: the low energy strength is concentrated into two major peaks  around 18~MeV and 22~MeV and a non negligible
 high energy component is found at 30 MeV.

In Table~\ref{tabM}, theoretical predictions of resonance mean energies are compared with available data \cite{YB2002,yb}.
Theoretical  values are calculated in the energy range [9~-~41~MeV] as was done for experimental data~\cite{YB2002,yb}.
Values obtained from the systematic law $\dspt E\left(GMR\right)$~=~80~$A^{-\frac{1}{3}}$~MeV are given in the same Table~\ref{tabM}.
\begin{table}
\caption{Theoretical (present work), systematic and experimental mean energy values of the GMR (in~MeV). }
%\begin{ruledtabular}
\begin{tabular}{cccc}
\hline\hline
&$E(GMR)$ &~MeV 
$$\\
&th.&Syst.& Exp.\\
\hline
$^{22}$Mg& 21.23 &28.5     &       \\
$^{24}$Mg& 21.06 & 27.7    &  21.0$\pm$0.6	\\
$^{26}$Mg& 21.99 & 27.0    &  	   \\
$^{28}$Mg& 21.33 &  26.3   &  	    \\
$^{26}$Si& 21.62 & 27.0    &  	  \\
$^{28}$Si& 21.76 &  26.3   &  21.25$\pm$0.38	   \\
$^{30}$Si& 21.85 &  25.7   &  	  \\
\end{tabular}\label{tabM}
%\end{ruledtabular}
\end{table}
Comparing our predictions with systematics and experimental data we find that: {\it i}) the mean energy is here predicted almost constant,
 irrespective of the mass of the nucleus, contrary to the $A^{-1/3}$ dependence of the systematic law, but
 {\it ii}) theoretical energies are smaller than systematics,
  QRPA results being close to experimental data in $^{24}$Mg and $^{28}$Si.

The splitting of monopole responses  for ASD nuclei is related to  the coupling with $K=0$ component of the quadrupole responses~\cite{ww}.

\subsection{Quadrupole response}\label{secQ}

\begin{figure*}
%\begin{center}
\centerline{\hbox{\psfig{file=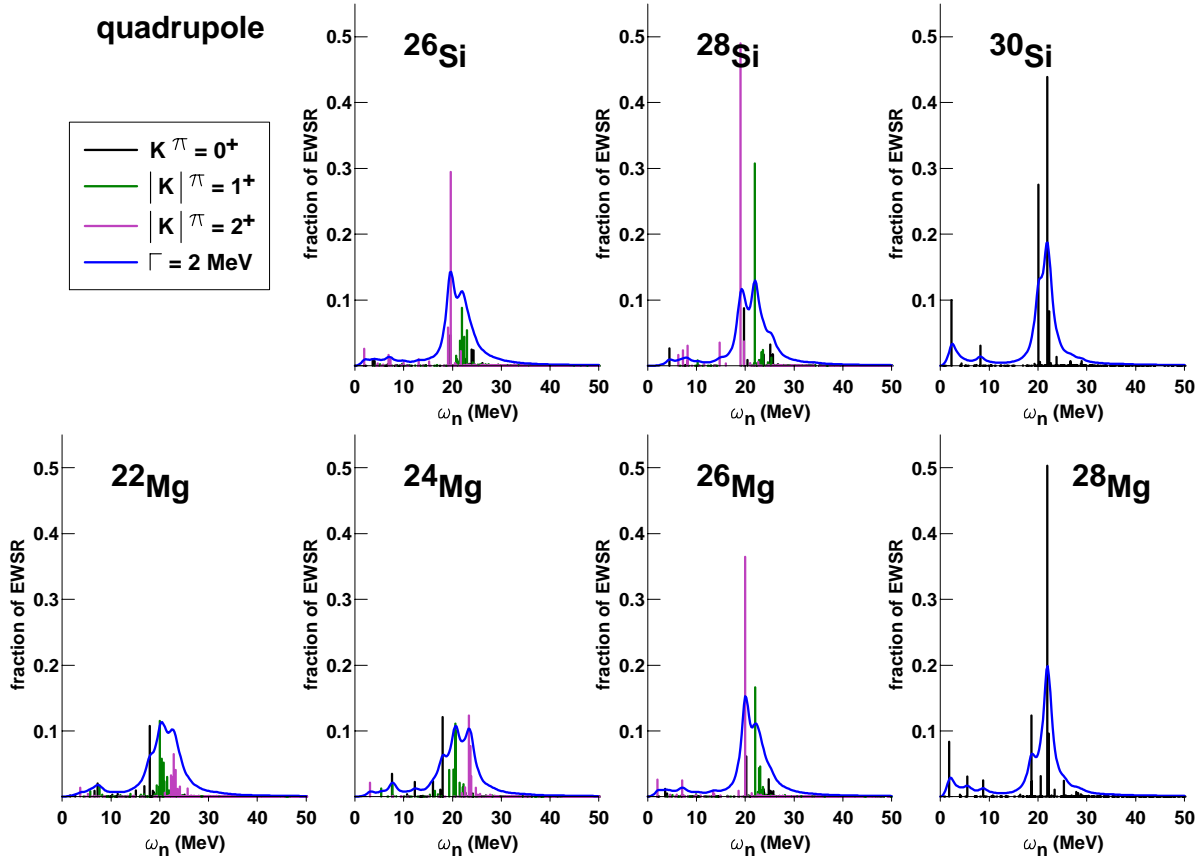,width=18cm,angle=0}}}
\caption{(Color online) Fractions of the quadrupole EWSR in $^{22-28}$Mg and in $^{26-30}$Si.
In the deformed nuclei, K$^\pi=0^+$, $\vert$K$\vert ^\pi=1^+$ and $\vert$K$\vert ^\pi=2^+$ components are indicated in black, green and purple, respectively.
 Blue curves are obtained from folding of QRPA spectra with a Lorentzian distribution.}
\label{fig6}
%\end{center}
\end{figure*}

Since the strength in light nuclei is found to be highly fragmented, 
the experimental identification of various giant resonances is difficult. 
As mentioned in references \cite{lisan,bertrand,cseh}, this led in the 70's 
to substantial disagreements between various studies~\cite{PRC13,PRC15,PRC76}.
QRPA results for quadrupole responses in the seven nuclei under study
 are displayed in Fig.~\ref{fig6}, where K$^\pi=0^+,\pm 1^+,\pm 2^+$ components are in black, green and  purple, respectively.
Discrete spectra have been folded with a Lorentzian distribution,
 $\dspt{\cal L}\left(x\right)={\Gamma} / {2\pi}\left(x^2+{\Gamma^2}/{4}\right)$ 
 with a $\Gamma$~=~2~MeV width, and the result plotted as blue lines. 

In all seven nuclei, quadrupole responses are predicted to lie between a few MeV and 30~MeV 
without any strength above, contrary to dipole and monopole responses which extend up to 50~MeV (see sections \ref{secD} and \ref{secM}).
Discrete low-lying states are found up to 18~MeV and a Giant Quadrupole Resonance (GQR) is predicted around 20~MeV.
Let us note that the first low-lying state is found to be $\vert$K$\vert^\pi=2^+$  
in all deformed nuclei except $^{28}$Si. 
The occurrence of $\vert$K$\vert ^\pi=2^+$ states at a very low energy is an indication of the softness of these s-d shell nuclei with respect to
the triaxial degree of freedom as mentioned in section \ref{sec3}.
As to monopole and dipole modes,
in the oblate nuclei ($^{26-28}$Si and $^{26}$Mg) $\vert$K$\vert$~=~J~=~2 components of the GQR have lower energies than
K~=~0 ones, while 
opposite behaviors are observed in prolate nuclei ($^{22-24}$Mg).
The energies  of the $K^{\pi}=1^+$ components are around  20~MeV in the two prolate Magnesium isotopes 
and close to 22~MeV in the oblate nuclei, that is in between the energies of the
 $K^\pi=0^+$ and of the $\vert$K$ \vert ^\pi$~=~$2^+$ components, in all deformed nuclei.
The same hierarchy was found by Hagino~\cite{hagino} et al in $^{24}$Mg and in $^{38}$Mg.
As the splitting of the quadrupole resonance due to the deformation is not very large, only one broad peak is found in the GQR.
The width of the GQR is found to be smaller than the ones of the giant dipole and monopole resonances.
However the energy splitting between K~=~0 and $\vert$K$\vert$~=~2 components is arround 5 MeV in  
agreement with the 12-14~A$^{-1/3}$MeV splitting obtained from macroscopic models (see section 4.7 of ref~\cite{ww}).
The quadrupole responses  of the two  spherical nuclei are not much  fragmented 
and are found to be more collective than dipole and monopole responses.

Table \ref{tabQ} gives the theoretical
mean energies of the GQR compared with  values from the systematics  E(GQR)~=~63~$A^{-\frac{1}{3}}$~MeV 
and available measurements \cite{yb,YB2002}.
The same energy range, [9~-~41~MeV], is considered to calculate the mean theoretical and experimental values~\cite{yb,YB2002}.
\begin{table}
\caption{theoretical (present work), systematic and experimental mean energy values of the GQR (in~MeV)}
%\begin{ruledtabular}
\begin{tabular}{cccc}
\hline\hline
&$E(GQR)$ &~MeV 
$$\\
&Th.&Syst.&Exp.\\
\hline
$^{22}$Mg& 20.62 & 22.5  &\\
$^{24}$Mg& 20.54 & 21.8  &16.9$\pm$0.6 \\
$^{26}$Mg& 20.97 & 21.3  &\\
$^{28}$Mg& 21.41 &  20.7 &\\
$^{26}$Si& 20.63 & 21.3  &\\
$^{28}$Si& 20.41 &  20.7 &18.54 $\pm$ 0.25 \\
$^{30}$Si& 21.44 &  20.3 &\\
\end{tabular}\label{tabQ}
%\end{ruledtabular}
\end{table}
Contrary to monopole results, theoretical GQR energies are close to systematic values 
but larger than experimental ones.
In the $^{24}$Mg nucleus, the theoretical energy of the GQR (exhausting  90$\%$ of the EWSR)
is 3.64~MeV larger than the  experimental one (exhausting only 72~$\pm$ $10\% $ of the EWSR)~\cite{yb}.
In the $^{28}$Si nucleus, the theoretical energy of the GQR (exhausting 91$\%$ of the EWSR) is 1.87~MeV larger than the
the experimental one (exhausting only 68 $\pm $ 9 $\%$ of the EWSR)~\cite{YB2002}.
The mismatch of 10$\%$ in $^{28}$Si and 20$\%$ in $^{24}$Mg with experimental values 
is probabely the consequence of the value of effective mass ($m^*/m=0.7$) of the D1S interaction 
which is the one giving correct single-particle properties in mean-field calculations.
Let us note that a reanalysis of the experimental data in $^{28}$Si nucleus has given a higher value of the 
centroid energy 18.77~$\pm$~0.35~MeV corresponding to 102~$\pm$~11~$\%$ of the E2 EWSR in the region [9~-~35~MeV]~\cite{YB2007}.
Keeping in mind that our theoretical description does not include any adjustable parameters,
the agreement with experimental data can be considered quite satisfactory.

\subsection{Octupole response}\label{secO}

\begin{figure*}
\begin{center}
\centerline{\hbox{\psfig{file=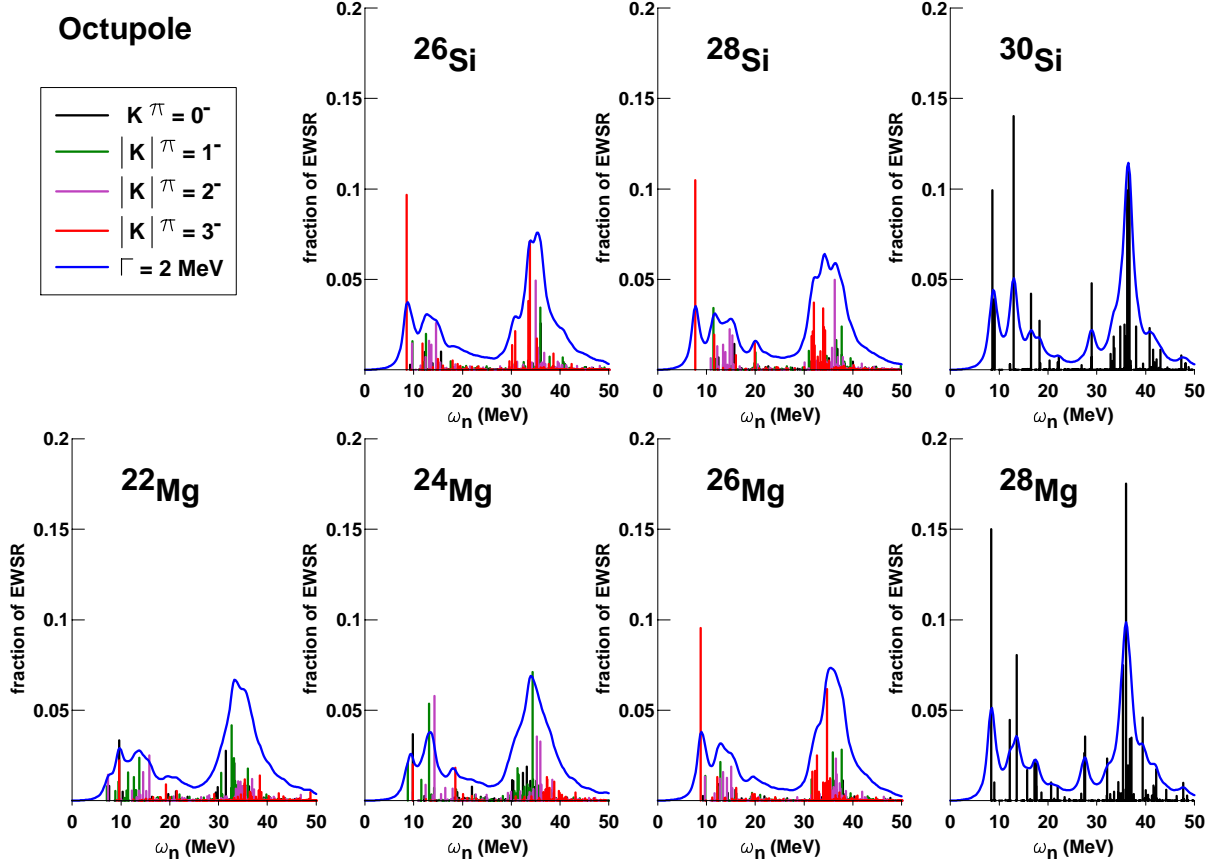,width=18cm,angle=0}}}
\caption{(Color online) Fractions of the octupole EWSR in  $^{22-28}$Mg and $^{26-30}$Si.
In the deformed nuclei, K$^\pi=0^-$, $\vert$K$\vert ^\pi=1^-$, $\vert$K$\vert ^\pi=2^-$ and $\vert$K$\vert ^\pi=3^-$
 components are indicated in black, green, 
purple and red, respectively. 
Blue curves are obtained from folding of QRPA spectra with a Lorentzian distribution.}
\label{fig7}
\end{center}
\end{figure*}

Fig. \ref{fig7} shows the octupole responses in $^{22-28}$Mg and  $^{26-30}$Si.
In this figure, K$^\pi=0^-$, K$^\pi=\pm 1^-$, K$^\pi=\pm 2^-$ and K$^\pi=\pm 3^-$ components are indicated 
in black, green, purple and red, respectively.
As for the quadrupole response, a folding of the QRPA spectra with a 2~MeV width Lorentzian distribution 
has been performed.
As known for a long time \cite{ww}, the octupole response is split into two energy components, the 
low energy octupole resonance (LEOR) generally collective and little fragmented, and the high energy octupole resonance (HEOR).
In all nuclei studied here, these two energy components are well identified,
the LEOR ranging from 5~MeV to 25~MeV and the  HEOR  from 25~MeV to 45~MeV.
In the oblate  $^{26-28}$Si and $^{26}$Mg, the LEOR is typically made of a highly collective 
$\vert$K$\vert^\pi=3^-$ state around 8~MeV and  K-mixed fragments  
 around 15~MeV. 
In the prolate $^{22-24}$Mg isotopes the LEOR  is even more fragmented without major contribution 
from any given angular momentum projection K.
In deformed nuclei, all K-components contribute to the HEOR, which is found to be highly fragmented.
However, the ordering of the different K-components is still found inverted between prolate and oblate nuclei:
between 30~MeV and 35~MeV 
$\vert$K$\vert ^\pi=1^-$ states are those contributing the most to the strength in prolate nuclei,
whereas in oblate nuclei, the strength is due mainly by $\vert$K$\vert ^\pi=3^-$ states.
Octupole responses in the spherical nuclei $^{30}$Si and $^{28}$Mg also show LEOR and HEOR components.

 Table \ref{tab3} gives the theoretical mean
 energies  of  LEOR and HEOR. They are calculated in the two intervals [~5~-~25~MeV~] and [~25~-~45~MeV~], respectively.
 Values calculated from the energy systematic law of the LEOR (E~=~30~$A^{-1/3}$~MeV) are also given.
\begin{table}
\caption{QRPA LEOR and HEOR mean energy values (in~MeV) obtained integrating from 5 to 25~MeV and 25 to 45~MeV, respectively.
 Values from the systematics of the LEOR are also given.}
%\begin{ruledtabular}
\begin{tabular}{ccccc}
\hline\hline
(MeV)&LEOR&&&HEOR\\
&Th.&Syst.&&Th.\\
\hline
$^{22}$Mg& 12.37 &10.70 && 34.48 \\
$^{24}$Mg& 13.49 &10.40 && 34.83  \\
$^{26}$Mg& 12.13 &10.12 && 35.72  \\
$^{28}$Mg& 11.18 & 9.87 && 35.23  \\
$^{26}$Si& 11.88 &10.12 && 34.97 \\
$^{28}$Si& 11.54 & 9.87 && 34.97 \\
$^{30}$Si& 11.67 & 9.65 && 35.63 \\
\end{tabular}\label{tab3}
%\end{ruledtabular}
\end{table}
The theoretical values of LEOR are larger than those from systematics. 
They rather follow a "corrected" systematics E~=~35~$A^{-1/3}$~MeV.
No systematics exists for HEOR in light nuclei.
Nonetheless our results are compatible with the systematic law E(HEOR)~=~110~A~$^{1/3}$~MeV, 
obtained from experimental data in heavier nuclei~\cite{ww}

\section{Conclusion}\label{sec4}

A fully consistent microscopic axially-symmetric-deformed QRPA approach has been developed and applied 
to light even-even s-d shell nuclei.
In these calculations, the same Gogny D1S effective force is used in both HFB and QRPA approaches in all ph, pp and hh channels.
Results on giant resonances have been obtained in $^{22-28}$Mg and $^{26-30}$Si isotopes and the influence of the ground state
intrinsic deformation on strength distributions has been discussed.
 
In deformed nuclei, theoretical isovector dipole responses are found split into two major components, as expected.
The splitting is found to be correlated with  specific families of ph transitions.
The strength is predicted to be almost equally distributed between the two energy components.
The two groups of transitions correspond to dipole vibrations along the major and the minor axes of the deformed nuclear systems.
K~=~0 components are found in the low energy part of the spectra in the prolate nuclei, and $\vert$K$\vert$~=~1 components in the oblate ones.

Isoscalar monopole resonances also display a splitting in deformed nuclei.
Such a splitting is not found when a spherical HFB state is used to construct QRPA solutions.
Isoscalar quadrupole and octupole resonances are found to be well fragmented in particular in well-deformed nuclei.
Such a fragmentation masks the deformation splitting and the different K broad components
are overlapping.
However, LEOR and HEOR components are predicted to be well separated,
and the ordering of the K components as a function of energy is found to be
related to the sign of the quadrupole moment.
Results obtained in $^{24}$Mg and in $^{28}$Si show that the Gogny interaction qualitatively reproduces
known resonances without resorting to any readjustment of parameters.

These results show that 
a fully microscopic QRPA approach of deformed nuclei is able to provide a wealth of structure information and a satisfactory
agreement with available data.
Extensions of the present work to other nuclei, in particular exotics ones and to phenomena such as pygmy states are under study.

\section*{Acknowledgments}
The authors want to thank Professor P.-F. Bortignon for the interest he has shown for this work.


\begin{thebibliography}{00}

\bibitem {thou61}	D.J. Thouless, Nucl. Phys. {\bf 22}, 78 (1961).	
\bibitem {RS} P. Ring and P. Schuck, {\it The Nuclear Many Body Problem} (Springer-Verlag, New-York, USA, 1980).
\bibitem {blaizot} J.-P. Blaizot and G. Ripka, {\it Quantum Theory of finite Systems} (MIT, Cambridge, MA, 1986).
\bibitem {shlomo1975} S. Shlomo and G. Bertsch, Nucl. Phys. {\bf A243}, 507 (1975).
\bibitem {bertsch1975} G. F. Bertsch and S. F. Tsai, Phys Rep. {\bf 18}, 125 (1975).
\bibitem {hama} I. Hammamoto, H. Sagawa, and X. Z. Zhang, Phys. Rev. C {\bf 57}, R1064 (1998); Nucl. Phys. {\bf A648}, 203 (1999).
\bibitem {hagino2001} K. Hagino, H. Sagawa, Nucl. Phys. {\bf A695}, 82 (2001).
\bibitem {grasso} M. Grasso,N. Sandulescu, Nguyen Van Giai, and R. J. Liotta, Phys. Rev. C {\bf 64}, 064321 (2001).
\bibitem {matsuo} M. Matsuo, Nucl. Phys. {\bf A696}, 371 (2001).
\bibitem {khan} E. Khan, N.Sandulescu, M. Grasso, and Nguyen Van Giai, Phys. Rev. C {\bf 66}, 024309 (2002). 
\bibitem {colo}  N. Paar, D. Vretenar, E. Khan and G. Colo\`o , Rep. Prog. Phys. {\bf 70}, 691 (2007).
\bibitem {kamerd93} S. Kamerdzhiev, J. Speth, G. Tertychny and V. Tselyaev, Nucl. Phys. {\bf A555}, 90 (1993).
\bibitem {colo2001} G. Colo, P.F. Bortignon, Nucl. Phys. {\bf A696}, 427 (2001);
                    D. Sarchi, P.F. Bortignon, G. Colò, Phys. Lett. {\bf B601}, 27 (2004).
\bibitem {kamerd2004} S. Kamerdzhiev, J. Speth and G. Tertychny, Phys. Rep. {\bf 393}, 1 (2004).
\bibitem {ger} M. Z. I. Gering and W. D. Heiss, Phys. Rev. C {\bf 29}, 1113 (1984).
\bibitem {yoshi} K. Yoshida, M. Yamagami, K. Matsuyanagi, Nucl. Phys. {\bf A779}, 99 (2006). 
\bibitem {hagino} K. Hagino, Nguyen Van Giai, H. Sagawa, Nucl. Phys. {\bf A731}, 264 (2004).  
\bibitem {rand} P. Moller and J. Randrup, Nucl. Phys. {\bf A514}, 49 (1990).
\bibitem {vret} D. Vretenar, T. Nik\v{s}i\'c, P. Ring, N. Paar, G.A. Lalazissis, and P. Finelli, Eur. Phys. J. {\bf A20}, 75 (2004).
\bibitem {dario} N. Paar, D. vretenar,  T. Nik\v{s}i\`c, P. Ring, Phys. Rev. C {\bf 74}, 037303 (2006).
\bibitem {litvi} E. Litvinova, P. Ring, D. Vretenar, Phys. Lett. {\bf B647}, 111 (2007).
\bibitem {fayans} S.A. Fayans, E.L. Trykov, D. Zawischa, Nucl. Phys. {\bf A568}, 523 (1994).
\bibitem {agra63} B.K. Agrawal, S. Shlomo, and A.I. Sanzhur, Phys. Rev. C {\bf 67}, 034314 (2003).
\bibitem {shlomo} B.K. Agrawal and S. Shlomo, Phys. Rev. C {\bf 70}, 014308 (2004).
\bibitem {epja} S. P\'eru, J.-F. Berger and P.-F. Bortignon, Eur. Phys. J. {\bf A26}, 25 (2005).
\bibitem {tera} J. Terasaki, J. Engel, M. Bender, J. Dobaczewski, W. Nazarewicz, and M. Stoitsov, Phys. Rev. C {\bf 71}, 034310 (2005).
\bibitem {tapas} Tapas Sil, S. Shlomo,B. K. Agrawal, and P.-G. Reinhard, Phys. Rev. C {\bf 73}, 034316 (2006).
%\bibitem {shlomo2003} B.K. Agrawal, S. Shlomo, and A.I. Sanzhur, Phys. Rev. C {\bf 67}, 034314 (2003).
\bibitem {tera2006} J. Terasaki and J. Engel, Phys. Rev. C {\bf 74}, 044301 (2006).
\bibitem {gog1}J. Decharg\'e and D. Gogny, Phys. Rev. C {\bf 21}, 1568 (1980).
               J.F. Berger, M. Girod, and D. Gogny, Comp. Phys. Comm. {\bf 63}, 365 (1991).
\bibitem {lisan} J. Lisantti, F.E. Bertrand, D.J. Horen, B.L. Burks, C.W. Glover, D.K. Mc Daniels, L.W. Swenson, X.Y. Chen, 
O. Hausser, K. Hicks, Phys. Rev. C {\bf37}, 2408 (1988).
\bibitem {bertrand} F. E. Bertrand, K. van der Borg, A. G. Drentje, M. N. Harakeh, J. van der Plicht, 
and A. van der Woude, Phys. Rev. Let. {\bf 40}, 635 (1978).
\bibitem {cseh} J. Cseh and I. Fodor, J. Phys G {\bf 11}, 103 (1985).
\bibitem {PRC13} D. H. Youngblood, J.M. Moss, C.M. Rozsa, J.D. Bronson, A.D. Bacher, D.R. Brown, Phys. Rev. C {\bf 13}, 994 (1976).
\bibitem {PRC15} D. H. Youngblood, C. M. Rozsa, J. M.Moss, D. R. Brown, and J. D. Bronson, Phys. Rev. C {\bf 15}, 1644 (1977).
\bibitem {PRC76} A. Kiss, C. Mayer-B\"oricke, M. Rogge, P. Turek, and S. Wiktor , Phys. Rev. Lett. {\bf 37}, 1188 (1976).
\bibitem {YB2002} D.H. Youngblood, Y.-W. Lui, and H.L. Clark, Phys. Rev. C {\bf 65}, 034302 (2002).
\bibitem {yb} D.H. Youngblood, Y.-W. Lui, and H.L. Clark, Phys. Rev. C {\bf 60}, 014304 (1999).	
\bibitem {russe} B.S. Ishkhanov, I.M. Kapitonov, E.V. Lazutin, V.G. Shevchenko, Nucl. Phys. {\bf A186}, 438 (1972).
\bibitem {russe2}K.M. Irgashev, B.S. Ishkhanov and I.M. Kapitonov, Nucl. Phys. {\bf A483}, 109 (1988).	  	
\bibitem {YB2007} D.H. Youngblood, Y.-W. Lui, and H.L. Clark, Phys. Rev. C {\bf 76}, 027304 (2007).
\bibitem {ww} M.N. Harakeh, A. van der Woude,  {\it Giant Resonances:
Fundamental High-energy Modes of Nuclear Excitation} (Oxford Un. Press, Oxford, 2001).
\bibitem {papa} P. Papakonstantinou, EPL, {\bf 78},  12001 (2007).
\bibitem {n2028} S. P\'eru, M. Girod, and J.F. Berger,  Eur. Phys. J. {\bf A9}, 35 (2000).
\bibitem {comex2} S. P\'eru, H. Goutte and J.-F. Berger, Nucl. Phys. {\bf A788}, 44c (2007).
\bibitem {BOT37} W. Bothe, and W. Gentner, Z. Phys. {\bf 71}, 236 (1937).
\bibitem {MIG44} A. Migdal, J. Phys {\bf 8}, 331 (1944).
\bibitem {BER75} B.L. Berman and S. C. Fultz, Rev. Mod. Phys. {\bf 47}, 713 (1975).
\bibitem {BER77b} R. Berg\`ere,  {\it Lecture Notes in Physics, {\bf 61} Photonuclear Reactions I} (Springer-Verlag, Berlin, 1977).
\bibitem {DIE88} F. S. Dietrich and B. L. Berman, At. Data Nucl. Data Tables {\bf 38}, 199 (1988).
\bibitem {oka} K. Okamoto, Phys. Rev. {\bf 110}, 143 (1958).
\bibitem {danos} M.Danos, Nucl. Phys. {\bf 5}, 23 (1958).
\bibitem {bass} W. H. Bassichis and F. Scheck, Phys. Rev. {\bf 145}, 771 (1966).
\bibitem {goriely} S. Goriely, E. Khan, Nucl. Phys. {\bf A706}, 217 (2002).

\end{thebibliography}
\end{document}